\documentclass[a4paper]{jpconf}

\usepackage{graphicx}
\usepackage{epstopdf}

\begin{document}
\title{Linking the Supermassive Black Hole Growth with the Megamaser Emission}

\author{Anca Constantin}

\address{James Madison University, Harrisonburg, VA 22807}

\ead{constaax@jmu.edu}

\begin{abstract}
High-resolution observations of the central few 100 pc of the galactic nuclear environments remain prohibitive for large statistical samples, which are crucial for tracing the links between central black hole formation, galaxy formation and AGN activity over cosmic time.  With this contribution, we present novel ways of connecting the physics of black hole accretion with its immediate environs via a new quantitative evaluation of the degree to which the strength and spatial configuration of the water maser emission is associated with the nuclear nebular galactic activity.  We discuss possible evolutionary/causal connections between these two types of emission, together with criteria that could dramatically enhance our search for mega-maser systems in nearby galactic centers.  These are timely results given the interest in combining high-resolution observations with extremely large optical telescopes and large arrays that start to conquer the sub-millimeter window.

\end{abstract}

\section{Introduction}

Mega-masers are remarkable natural phenomena.  Mapping of the H$_2$O maser emission  with $L_{\rm H_2O} \geq 10 L_{\odot}$ in a disk configuration has provided the first direct evidence in an active galactic nucleus (AGN) for the existence of a thin Keplerian accretion disk with turbulence \cite{greenhill95}, highly compelling evidence for the existence of a massive black hole (BH) \cite{miyoshi95, kuo11}, as well as a cosmic distance determination of extremely high precision \cite{herrnstein99,herrnstein05}.  To date, such observations and associated calculations have been performed on only less than a handful such systems \cite{reid09,kuo11}. Identifying and studying a large sample of the mega-masers in this particularly rare disk configuration is of crucial relevance for our progress in understanding the accretion processes in galaxy centers, and more importantly, for constraining the proposed cosmological models and the nature of Dark Energy.   Unfortunately, good candidates are extremely few; the success rate in maser detection remains a mere $\sim 3-5\%$, out of which only $\sim 40\%$ appear to originate in disks \cite{braatz96, braatz97, braatz04, greenhill03, kondratko06, braatz10}.

To efficiently search for new powerful mega-maser disks, we need a good understanding of the special physical characteristics that facilitate their production in galaxy centers.  There is some evidence that mega-masers may be associated with the molecular disk or torus that surrounds and (partially) obscures an actively accreting massive black hole ($M_{\rm bh} \sim 10^7 M_{\rm sun}$) harbored by a galactic nucleus, i.e., an AGN \cite{miyoshi95, braatz04}.    However, mega-masers are also detected in face-on, unobscured disks or elliptical galaxies where no molecular torus is expected, and even along jets protruding long distances from the nucleus \cite{falcke00, ferruit00}.  So, either masers are not always produced in association with an AGN torus, or these particular cases are unrecognized AGN.  Indeed, for the majority of the nearby active galaxies, an AGN classification remains ambiguous; in these sources, an actively accreting black hole might be either obscured by large amounts of dust in the surrounding star-forming regions, or simply hidden in a mix of other ionization sources (shocks, turbulence, etc.) or host galaxy light \cite{ho2008}.  Thus, if mega-maser production is genuinely related to black hole accretion, the true detection rate of disk mega-maser emission remains unknown as long as the AGN census is incomplete. To shed light on these matters, we need to investigate all possible connections between the H$_2$O emission and their host properties, for a wide variety of galactic nuclear activity.  We only now have the number statistics necessary to perform such a study.

We introduce here an in-depth multi-parameter characterization of the optical properties of the largest available sample of 146 galaxy centers hosting H$_2$O masers together with a $\sim 20$ times larger control sample of galaxies in which maser detection failed.   This effort complements the very recent work by Zhu et al. \cite{zhu11}, where they investigate the optical properties of the maser host galaxies and find that the maser detection rate increases for higher optical luminosity, larger velocity dispersion, higher extinction and higher [O III] $\lambda$5007 line luminosity.     Our analysis adds a novel comparative investigation of the optical host and nuclear features separated per spectral types that correlate with the strength and type of nuclear galactic activity.     If, as previously suggested \cite{constantin08, constantin09, schawinski07, schawinski10}, there is an evolutionary sequence in which galaxies transform from star-forming via AGN to quiescence, accretion onto the central BH should happen as early as the star formation (or H {\sc ii}) phase, and should be active in Transition Objects (or $T$s).   Thus, if maser activity is related to accretion, we should look for (mega)masers in H {\sc ii}'s and Ts as well.   Restricting our searches to Seyferts ($S$s) and Low Ionization Nuclear Emission Regions (or LINERs; $L$s), as it appears to be the case with current efforts, could overlook a good portion, or even the lion's share of mega-masers.


\section{The Sample Galaxies \& Parameters Involved}

The Mega-maser Cosmology Project (MCP; \cite{reid09, braatz10}) makes publicly available the entire sample of galaxies detected in H$_2$O by May 2011. The data is presented in the form of an atlas that comprises sky positions, recession velocities, the maser spectra, and the corresponding discovery reference, for 146 such systems.  The MCP atlas also includes the list of all $>3300$ galaxies that have been surveyed for H$_2$O maser emission, along with their Green Bank Telescope (GBT) spectra, recession velocities, the sensitivity of each observation, and the corresponding source brightness temperature.  These data provide for the first time a sufficiently large sample of maser galaxies and a control sample of non-detections whose properties can be compared at a statistically significant level.

Optical spectral and photometric measurements for these objects are obtained mainly from wide sky surveys;  for the purpose of the analysis presented here, we exploited the nearby galaxy sample of the seventh data release of the SDSS \cite{abazajian09}.  Measurements of the host and nuclear nebular emission of the SDSS galaxies are drawn from the catalog built by the MPA/JHU collaboration \cite{brinchmann04}.  For these objects, the emission-line component is measured after it is separated from the host stellar emission, and subtracted from the total galaxy spectrum based on fits of stellar population synthesis templates.  Black Hole masses ($M_{\rm BH}$) and corresponding accretion rates $dM/dt$ (measured via the Eddington ratio $L/L_{\rm Edd}$) are obtained based on estimates of the stellar velocity dispersion ($\sigma^*$), that are available via careful spectral fitting of the host stellar light component in galaxy spectra.  To relate the maser and the central BH accretion activity to the host properties, we employ stellar masses and the $D_n4000$ break index 
or H$\delta _A$ Balmer absorption-line index as proxies for the age of the associated stellar population, as calculated and presented by Kauffman et al. \cite{kauffmann03,kauffmann04}.

Cross-matches of the maser and control galaxy lists with these SDSS-based catalogs uncover 
79 maser and $\sim 1600$ control galaxies  with photometric information, among which high quality  spectroscopic measurements ($> 2 \sigma$ confidence) are available for 46 and 1181 maser and non-maser systems respectively.  We present here only a comparison between the spectroscopic characteristics.
It may be important to note that considering only the spectroscopic samples the 3\% maser detection rate remains in place.


\section {A statistical comparison of the properties of maser and non-maser galaxies}

\begin{figure}[h]
\hspace{-1.5pc}\begin{minipage}{26pc}
\includegraphics[width=29pc]{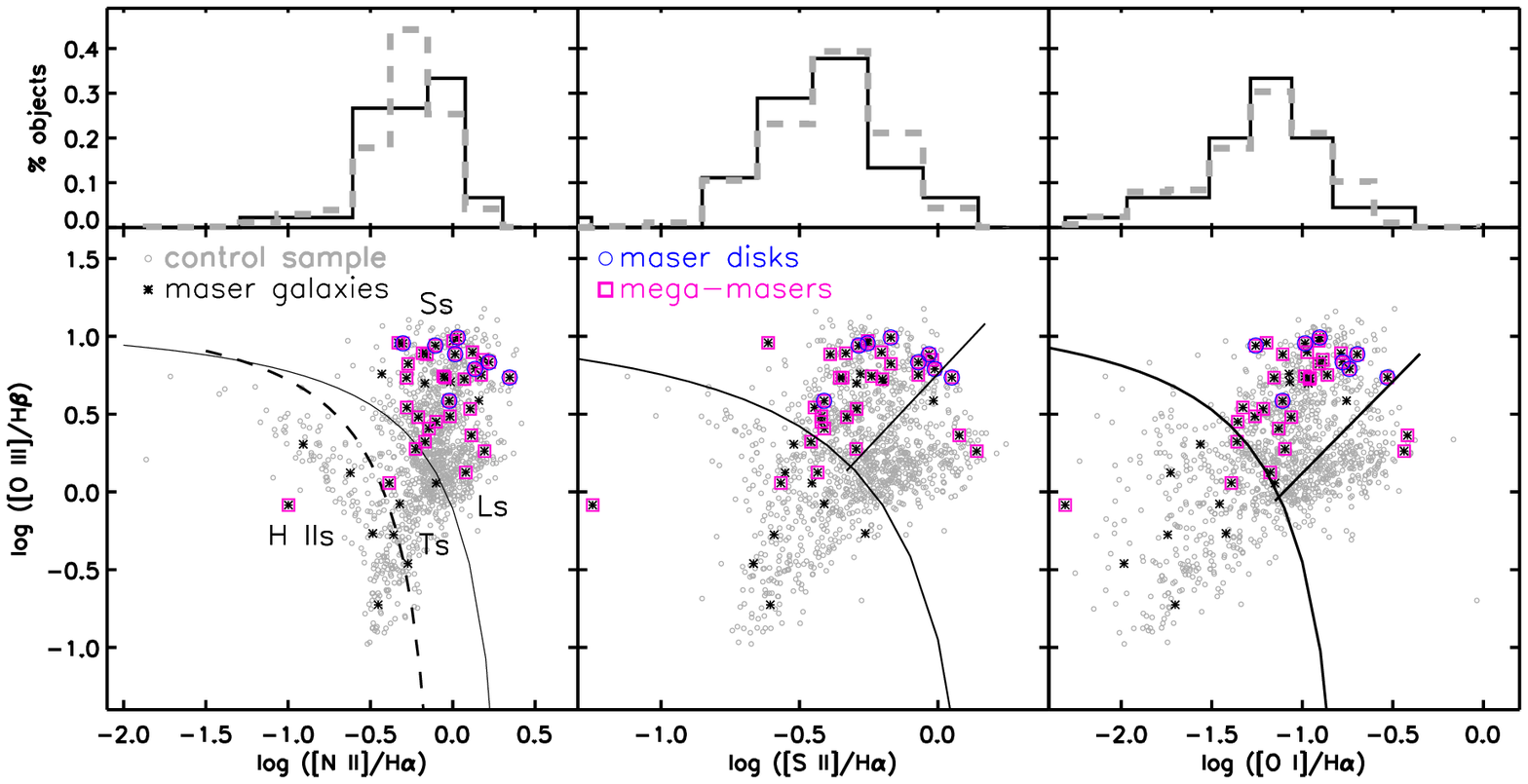}
\caption{\small \label{bpt} Line-diagnostic diagram for all maser and control galaxies, with a comparison of histograms of the involved line ratios.  The non-maser galaxies are plotted with grey symbols, the masers in black, the mega-masers ($L_{\rm H_2O} \geq 10 L_{\rm sun}$) are shown as magenta squares while the disks among them are the blue circles.  The lines in the bottom panels are semi-empirical fits to the distributions of objects in these diagrams.}
\end{minipage}\hspace{1.5pc}%
\begin{minipage}{13pc}
\includegraphics[width=13.5pc]{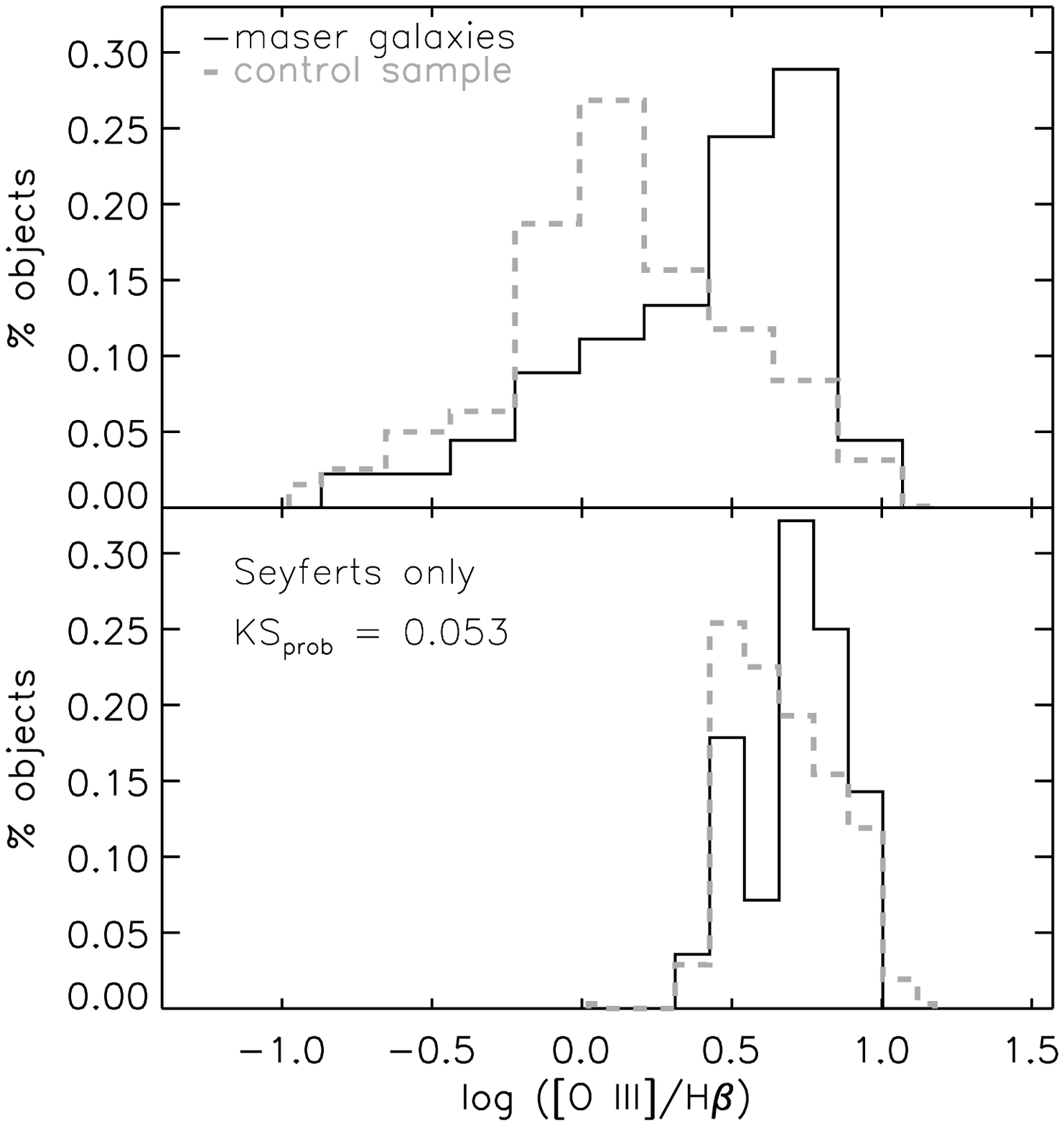}
\caption{\small \label{hist_o3hb} Comparison of the distributions of [O III]/H$\beta$ line ratios for maser and control galaxies.  [O~III]/H$\beta$ is clearly higher in maser hosts. }
\end{minipage} 
\end{figure}

Optical emission line ratios have often proven to be useful in identifying the dominant ionization mechanism or in delineating differences between various types of galaxy nuclei \cite{bpt81, veilleux87, kewley06}.    A comparison of the emission-line activity in maser and non-maser galaxies is presented in Figures ~\ref{bpt} and ~\ref{hist_o3hb} where we also show how the distributions of the line ratios compare for the two samples.   To classify the galaxies into $S$s, $L$s, $T$s, and H {\sc ii}s, we adopt the criteria of Kewley et al. \cite{kewley06}.

With the exception of the [O~III/H$\beta$] ratio, which is clearly larger for the maser sample, there is no obvious statistical difference between the maser and non-maser (control) galaxies.    The difference in [O~III/H$\beta$] could simply be a consequence of the fact that the bulk of the masers have been detected in surveys of Seyferts, whose definition requires high such ratios; however, when only Seyfert maser and control galaxies are compared, the difference is still statistically significant (KS$_{\rm prob}$ = 0.053), implying that it reflects an intrinsic relation to maser activity.  

The maser detection rate among Seyferts is high ($\sim 8\%$), however, it is unclear whether this is because the disk-maser emission is genuinely connected to this type of nuclear activity (i.e., the BH accretion process) or is simply a consequence of the same survey bias.
The maser disks are usually mega-masers, but not all mega-masers are disks. The disks have only been found among the Seyferts, however, mega-masers appear to cover the whole spectral spectrum.  The maser detection rate is clearly non-zero for non-AGN types, in fact, there is an equal probability to find masers among $T$s as in $L$s ($\sim$1.6\%), and twice as likely to find them in H {\sc ii}s ($\sim 3\%$).   

\subsection{The host characteristics that differentiate between maser and non-maser galaxies}

\begin{figure}[h]
\hspace{-1.95pc}
\includegraphics[width=44pc]{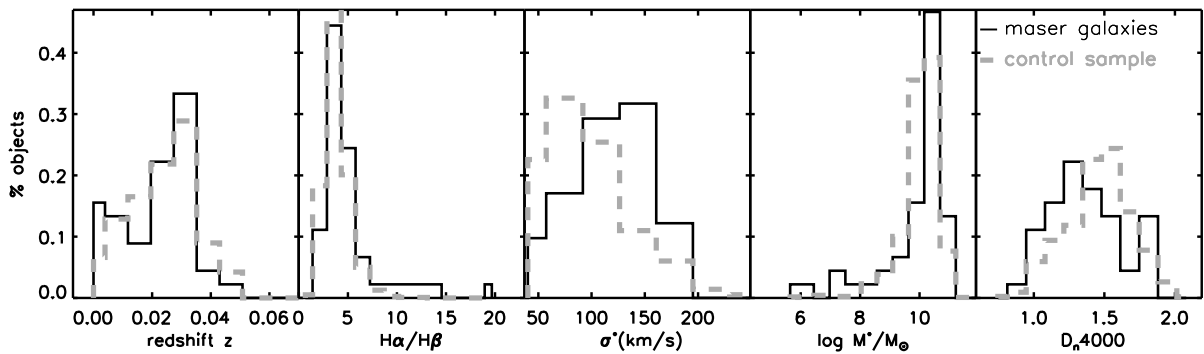}
\begin{minipage}{38pc}\caption{\small \label{hist_host} Distributions of redshifts $z$, Balmer Decrements $H\alpha/H\beta$, stellar velocity dispersions $\sigma^*$, host stellar mass $M^*$, and $D_n4000$ break index for maser galaxies (black lines) and the control sample (dashed gray lines).}
\end{minipage}
\end{figure}

Our comparisons of overall distributions of a variety of optical host properties of maser and non-maser galaxies support previous findings that: 1) maser galaxies show statistically higher levels of reddening and extinction, as measured by the H$\alpha$/H$\beta$ Balmer decrements; only the maser galaxies show values of H$\alpha$/H$\beta$ $>10$;  2) maser galaxies are more massive (higher stellar mass $M^*$), more luminous (in $M_r$, not shown here), and harbor more massive BHs in their centers (as shown by the comparison in $\sigma^*$).   We also show here for the first time that there is a possible connection between the maser activity and the age of the stellar population associated with their host galactic nuclei; the maser galaxies exhibit younger stellar populations (weaker $D_n4000$, suggesting an order of magnitude difference in age, in average).   Figure ~\ref{hist_host} presents these results.  

\begin{figure}[h]
\hspace{-1.5pc}\begin{minipage}{18pc}
\includegraphics[width=20pc]{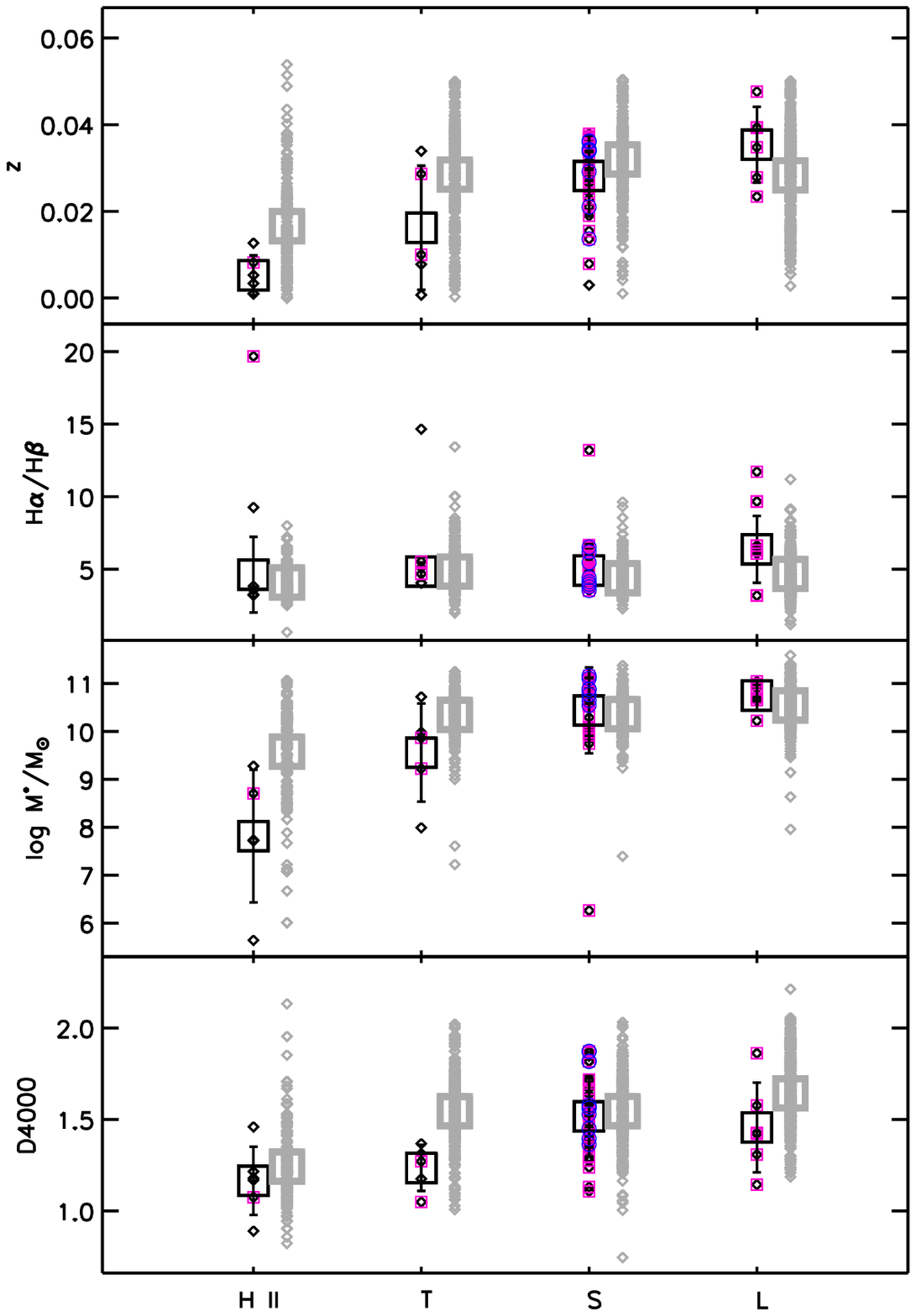}
\caption{\small \label{sequence_host} Host properties compared as a function of the different optical spectral types H~{\sc ii}, T, S, and L.   Symbol types are as in Figure ~\ref{bpt}.  Large open squares indicate averages per spectral type.}
\end{minipage}\hspace{1.0pc}%
\begin{minipage}{18pc}
\includegraphics[width=20pc]{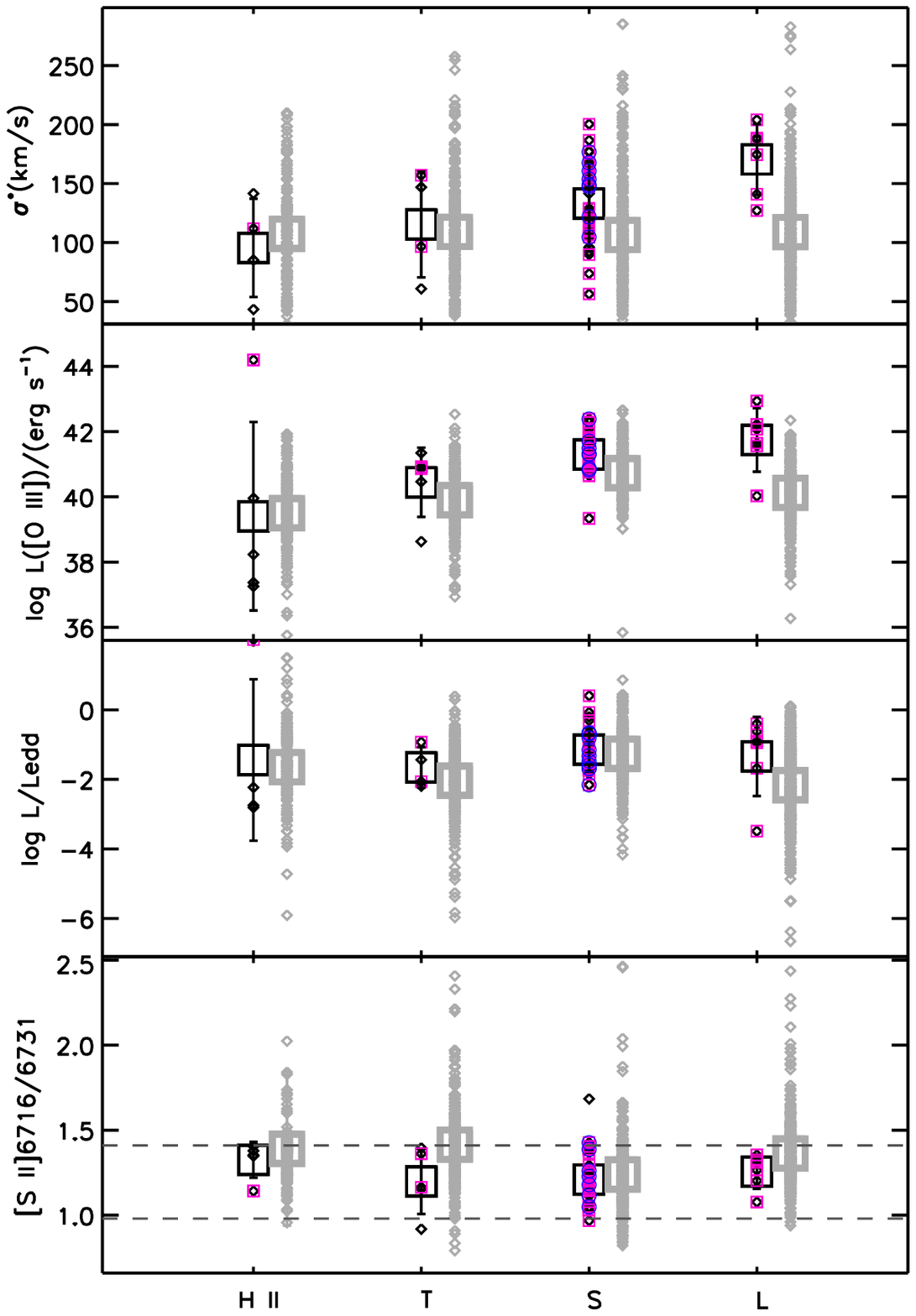}
\caption{\small \label{sequence_nuc} Nuclear properties compared as a function of  the different optical spectral types H~{\sc ii}, T, S, and L.   Symbol types are as in Figure ~\ref{bpt}.  Large open squares indicate averages per spectral type.}
\end{minipage} 
\end{figure}

Given the apparent bias of the current maser surveys toward Seyfert galaxies, interpreting  these global trends might benefit from looking into how these properties compare when maser and non-maser galaxies are separated per spectral types, as defined based on the line-diagnostic diagrams (Figure ~\ref{bpt}).   Figure ~\ref{sequence_host} illustrates the individual measurements of redshift, Balmer decrements, stellar masses, and $D_n4000$ parameters, for maser and non-maser galaxies, separated by their respective spectral types H {\sc ii}, T, S, and L.   This comparison reveals some features worth mentioning:  i)  H {\sc ii}'s with masers are only at very low $z$, and only if they are dwarfish, i.e., with very low stellar masses; ii) Ts associated with maser emission are more nearby, less massive and clearly younger in their stellar populations than their counterparts without maser activity; iii) interestingly, Ss with and without masers don't separate much from each other in these host properties; iv) it is only the high-$z$ Ls with the youngest stellar populations that host masers, which are all mega-masers.   Interestingly, the mega-maser disks are not among the galaxies with the youngest stellar populations, or among those with extremely large Balmer decrements.  It is probably surprising that the only H {\sc ii} hosting mega-maser emission has the highest Balmer decrement in the whole maser galaxy sample; nevertheless, this is not a mega-maser disk. 

\subsection{The nuclear features that differentiate between maser and non-maser galaxies}

A very similar type of comparison for measures that pertain more directly to the process of BH growth in these systems is presented in Figure ~\ref{sequence_nuc}.    Here we show individual measurements in 1) stellar velocity dispersion $\sigma^*$, as a measure of the BH mass;  2) [O III] line luminosity, $L{\rm [O III]}$, which is corrected for reddening and extinction based on a $\tau \propto \lambda^{-0.7}$ attenuation law \cite{charlot00} and the corresponding Balmer decrements; 3)  a measure of the Eddington ratio $L/L_{\rm edd}$,  where $L = L_{\rm bol} = 600 \times L{\rm [O III]}$ was used for the bolometric correction \cite{heckman04, kauffmann09} and $L_{\rm edd}$ is the Eddington luminosity corresponding to the $M_{\rm BH}$ values obtained via $M_{\rm BH} -\sigma^*$ relation of Graham et al. \cite{graham11}; and 4) the [S II] $\lambda$ 6716/$\lambda$6731 line ratio as a proxy for the density $n_e$ of the line-emitting gas \cite{osterbrock89}, which may also be viewed as  a gauge of the availability of fuel for accretion and/or masing activity.

We find here that the previously claimed trends for higher BH masses and $L{\rm [O III]}$ in maser galaxies \cite{zhu11} are correct not only for the Seyfert hosts, but for galaxies of all spectral types.  
While it seems especially true that among Ls, maser activity is detected only in systems with the highest $L{\rm [O III]}$ and the most massive BHs, Seyferts with and without masers exhibit a relatively weak statistical trend in this direction.    
Note also that galaxies with very massive BHs ($\sigma^* > 200$ km/s) do not host masers; physical conditions in the vicinity of such systems may not be  stable enough to sustain strong mega-maser emission.

Additionally, our comparison, reveals some new and potentially very useful constraints that could be posed to future maser surveys, particularly when searching for mega-maser disks: 

1) the combination of $M_{\rm BH}$ and $L{\rm [O III]}$ used to calculate $L/L_{\rm edd}$ exposes a potentially narrow range in the accretion rate of nuclei associated with maser emission, which in average is higher than in the control sample, for all the different spectral types compared here; it seems that the maser activity, and especially the mega-maser disk-like emission, is associated to accretion onto BHs characterized by $log L/L_{\rm edd} \sim -2$.   Incidnetally, this value also appears to correspond to a change in the accretion flow geometry or mode, from efficient (i.e., Shakura-Sunyaev standard accretion disk/corona, or quasar-like, \cite{shakura73}) to inefficient (i.e., the radiatively inefficient, or advection dominated accretion flow; RIAF/ADAF; e.g., \cite{narayan94}) with decreasing $L/L_{\rm edd}$ \cite{constantin09, gu09}.    It may thus be possible that the disk mega-maser emission is related to a certain (short) phase in the galaxy-AGN evolution, reflected by this particular change. 

2) the comparison of the [S II] $\lambda$ 6716/$\lambda$6731 line ratio unfolds a potential "goldilocks range" in $n_e$ for the maser galaxies, spanning only a range of $\sim 100 - 300$ cm$^{-3}$  (corresponding to a range in [S II] $\lambda$ 6716/$\lambda$6731 $= 1.4 - 1$).   The galaxies with the highest and lowest $n_e$ in the surveyed sample do not host maser activity at the levels that can be currently detected.   Because these measurements of $n_e$ reflect physical conditions of regions located way outside the masing areas, this parameter relates to the maser detection rate only indirectly; nevertheless, this connection is not less essential than that between $L{\rm [O III]}$ and the maser activity, which suffers from similar caveats.

\section{Summary}

We presented here some results of the first multi-parameter comparative analysis of the largest sample available of SDSS-based galaxies with and without maser activity in their centers. 
Besides confirming previously claimed tendency for high $L{\rm [O III]}$ in maser galaxies, we found new 
criteria for more efficient targeting of maser galaxies, and more importantly, of hosts of mega-maser disks, based on their optical spectral properties.    Our study suggests that mega-masers could be found in emission-line galaxies that are not necessarily classified spectroscopically as Seyferts, and that more appropriate survey designs should consider, at least to start with, a cross-match of narrow ranges in their $L/L_{\rm edd}$ values and the [S II] $\lambda$ 6716/$\lambda$6731 line ratios, as described above.    Considering a survey of all galaxies included in this study with $-2 < log L/L_{\rm edd} < 0$ and [S II] $\lambda$ 6716/$\lambda$6731 $= 1 - 1.4$, the maser detection rate is boosted by a factor of 4, to $\sim 12\%$, while the maser disk detection rate increases to $> 25\%$.   
More details on this analysis, along with a comprehensive Principal Analysis that encompasses a wider variety of parameters than those included in this discussion will be available in Constantin et al \cite{constantin12}.


\section*{References}
\begin{thebibliography}{99}

\bibitem{abazajian09} Abazajian, K.N., Strauss, M.A., et al., 2009, {\it ApJS}, {\bf 182}, 543

\bibitem {bpt81} Baldwin, J. A., Phillips, M. M., \& Terlevich, R. 1981, {\it PASP}, {\bf 93}, 5
\bibitem{braatz96}  Braatz, J. A., Wilson, A. S., \& Henkel, C. 1996, {\it ApJS}, {\bf 106}, 51
\bibitem{braatz97} Braatz, J. A., Wilson, A. S., \& Henkel, C. 1997, {\it ApJS}, {\bf 110}, 321
\bibitem{braatz04} Braatz, J. A., Henkel, C., Greenhill, L., Moran, J. \& Wilson, A., 2004, {\it ApJ}, {\bf 617}, L29
\bibitem{braatz10} Braatz, J. A., Reid, M. J., Humphreys, E. M. L., Henkel, C., Condon, J. J., \& Lo, K. Y., 2010, {\it ApJ}, {\bf 718}, 657
\bibitem{brinchmann04} Brinchmann, J., Charlot, S., Heckman, T. M., Kauffmann, G., Tremonti, C., \& White, S. D. M. 2004, arXiv:astro-ph/0406220
\bibitem{charlot00} Charlot, S., \& Fall, S. M. 2000, {\it ApJ}, {\bf 539}, 718
\bibitem{constantin08} Constantin, A., Hoyle, F., \& Vogeley, M. S., 2008, {\it ApJ}, {\bf 673}, 715
\bibitem{constantin09} Constantin et al. 2009, {\it ApJ},  {\bf 705}, 1336
\bibitem{constantin12} Constantin et al. 2012, in preparation.
\bibitem{falcke00} Falcke, H., Nagar, N.M., Wilson, A. S., \& Ulvestad, J.S., 2000, {\it ApJ}, {\bf 542}, 197
\bibitem{ferruit00} Ferruit, P., Wilson, A. S. \& Mulchaey, J., 2000, {\it ApJ}, {\bf 128}, 139
\bibitem{graham11} Graham, A.W., Onken, C.A., Athanassoula, E., \& Combes, F.,  2011, {\it MNRAS}, {\bf 412}, 2211 
\bibitem {greenhill95} Greenhill, L.J., Henkel, C., Becker, R., Wilson, T.L. \& Wouterloot, J.G.A., 1995, {\it A\&A}, {\bf 304}, 21
\bibitem{greenhill03} Greenhill, L. J., Kondratko, P. T., Lovell, J. E. J., Kuiper, T. B. H., Moran, J. M., Jauncey, D. L., \& Baines, G. P. 2003, {\it ApJ}, {\bf 582}, L11
\bibitem{gu09} Gu, M., \& Cao, X. 2009, {\it MNRAS}, {\bf 399}, 349
\bibitem{heckman04} Heckman, T. M., Kauffmann, G., Brinchmann, J., Charlot, S., Tremonti, C., \& White, S. D. M. 2004, {\it ApJ}, {\bf 613}, 109
\bibitem {herrnstein99} Herrnstein, J. et al., 1999, {\it Nature}, {\bf 400}, 539
\bibitem {herrnstein05} Herrnstein, J., et al. 2005, {\it ApJ}, {\bf 629}, 719
\bibitem{ho2008} Ho, L.C., 2008, {\it ARA\&A}, {\bf 46}, 475
\bibitem{kauffmann03} Kauffmann, G., et al. 2003, {\it MNRAS}, {\bf 341}, 33 
\bibitem{kauffmann04} Kauffmann, G., et al. 2004, {\it MNRAS}, {\bf 353}, 713 
\bibitem{kauffmann09} Kauffmann, G., \& Heckman, T. M. 2009, {\it MNRAS}, {\bf 397}, 135
\bibitem{kewley06} Kewley, L.J., Groves, B., Kauffmann, G., \& Heckman, T., 2006, {\it MNRAS}, {\bf 372}, 961
\bibitem{kondratko06} Kondratko, P. T. et al. 2006, {\it ApJ}, {\bf 638}, 100
\bibitem{kuo11} Kuo, C. Y., Braatz, J. A., Condon, J. J., Impellizzeri, C. M. V., Lo, K. Y.; Zaw, I., Schenker, M., Henkel, C., Reid, M. J., \& Greene, J. E., 2011, {\it ApJ}, {\bf 727}, 20
\bibitem {miyoshi95} Miyoshi, M., Moran J.M., Herrnstein J., Greenhill L., Nakai N., et al.1995, {\it Nature}, {\bf 373}, 127
\bibitem {narayan94} Narayan, R., \& Yi, I. 1994, {\it ApJ}, {\bf 428}, 13
\bibitem{osterbrock89} Osterbrock, D. E. 1989, {\it Astrophysics of Gaseous Nebulae and Active Galactic
Nuclei} (Mill Valley: University Science Books)
\bibitem {reid09} Reid, M. J., Braatz, J. A., Condon, J. J., Greenhill, L. J., Henkel, C., \& Lo, K. Y. 2009, {\it ApJ}, {\bf 695}, 287
\bibitem{schawinski07} Schawinski, K., et al. 2007, {\it MNRAS}, {\bf 382}, 1415
\bibitem{schawinski10} Schawinski et al. 2010, {\it ApJ}, {\bf 714}, 108
\bibitem{shakura73} Shakura, N. I., \& Syunyaev, R. A. 1973, {\it A\&A}, {\bf 24}, 337
\bibitem{veilleux87} Veilleux, S., \& Osterbrock, D. E. 1987, {\it ApJS}, {\bf 63}, 295
\bibitem{zhu11} Zhu, G., Zaw, I., Blanton, M. R., \& Greenhill, L. J., 2011, {\it ApJ}, {\bf 742}, 73


\end{thebibliography}
\end{document}